%% file: gSHEL_TiltedPolarizingInterface.tex
\documentclass[twocolumn,fleqn]{svjour3}

\def\makeheadbox{{}}

\input{preamble}
\input{commands}

\usepackage{url}

\begin{document}

\title{Geometric Spin Hall Effect of Light at Polarizing Interfaces}
\author{
Jan Korger
\and
Andrea Aiello
\and
Christian Gabriel
\and
Peter Banzer
\and
Tobias Kolb
\and
Christoph Marquardt
\and
Gerd Leuchs}

\institute{
Jan Korger
\and
Andrea Aiello
\and
Christian Gabriel
\and
Peter Banzer
\and
Tobias Kolb
\and
Christoph Marquardt
\and
Gerd Leuchs
\at
Max Planck Institute for the Science of Light,
Guenther-Scharowsky-Str. 1/Bau 24, 
91058 Erlangen,
Germany\\
Tel: +49 9131 6877 125\\
Fax: +49 9131 6877 199\\
\email{jan.korger@mpl.mpg.de}
\and
Jan Korger
\and
Andrea Aiello
\and
Christian Gabriel
\and
Peter Banzer
\and
Tobias Kolb
\and
Christoph Marquardt
\and
Gerd Leuchs\at
Institute of Optics, Information and Photonics,
University Erlangen-Nuremberg,
Staudtstr. 7/B2,
91058 Erlangen,
Germany
}

\date{2011-02-07}

\maketitle

\input{abstract}

\input{intro}

\input{TiltedField}
\input{TiltedPol}
\input{TiltedPolExp}
\input{PolShift}
\input{conclusion}

\input{ack}

The final publication is available at
\url{www.springerlink.com}.

\bibliographystyle{spphys}
\bibliography{Korger_ApplPhysB}

\end{document}

%% file: preamble.tex
\usepackage{graphicx}

\usepackage{fix-cm}	
\usepackage{mathptmx}   

\usepackage{amssymb}
\usepackage{amsmath} 
\usepackage[amssymb,textstyle,thinspace,thinqspace]{SIunits}

\usepackage[utf8]{inputenc}

\journalname{Applied Physics B}

%% file: commands.tex
\newcommand{\vh}[1]{\vec{\hat{#1}}}
\newcommand{\suprm}[1]{\ensuremath{^{\textrm{#1}}}}
\newcommand{\subrm}[1]{\ensuremath{_{\textrm{#1}}}}
\newcommand{\bvk}{(\vec k)}
\newcommand{\bvr}{(\vec r)}
\newcommand{\bvrp}{(\vrp)}
\newcommand{\vrp}{\vec r_\perp}
\newcommand{\bt}{(\theta)}

\newcommand{\tx}{\ensuremath{\tilde x}}
\newcommand{\ty}{\ensuremath{\tilde y}}
\newcommand{\tz}{\ensuremath{\tilde z}}
\newcommand{\bxez}{(\tx, \ty, \tz)}

\newcommand{\dif}{\,\mathrm{d}}
\newcommand{\dxy}{\dif x \dif y}
\newcommand{\Sum}{\sum\limits}
\newcommand{\eP}{\ensuremath{\vh e_P}}

\newcommand{\lrangle}[1]{\langle {#1} \rangle}

\newcommand{\EDbarycenter}[1][TD]{\ensuremath{\lrangle{y\subrm{#1}}\suprm{ED}}}

\newcommand{\Sbarycenter}[1][TD]{\ensuremath{\lrangle{y\subrm{#1}}^{\vec S}}}

\newcommand{\gO}{\mathcal O}

\newcommand{\pol}{po\-lar\-i\-za\-tion}
\newcommand{\poldep}{\pol-de\-pen\-dent}

\newcommand{\dimensional}{di\-men\-sion\-al}
\newcommand{\dep}{de\-pen\-dent}

\newcommand{\secref}[1]{section \ref{sec:#1}}
\newcommand{\figref}[2][]{Fig.\ \ref{fig:#2}#1}

\newcommand{\AAetal}{Aiello et al.}

%% file: abstract.tex
\begin{abstract}

The geometric Spin Hall Effect of Light (geometric SHEL)
amounts to a \poldep{} positional shift when a light beam
is observed from a reference frame
tilted with respect to its direction of propagation.
Motivated by this intriguing phenomenon,
the energy density of the light beam is decomposed into its Cartesian components
in the tilted reference frame.
This illustrates the occurrence of the characteristic shift and
the significance of the effective response function of the detector.

We introduce the concept of a tilted polarizing interface and provide a scheme for its experimental implementation.
A light beam passing through such an interface undergoes a shift resembling the original geometric SHEL in a tilted reference frame.
This displacement is generated at the polarizer and its
occurrence does not depend on the properties of the detection system.
We give explicit results for this novel type of geometric SHEL
and show that at
grazing incidence
this effect amounts to a displacement of multiple wavelengths, a shift larger
than the one introduced by
Goos-Hänchen and Imbert-Fedorov effects.

\PACS{
42.25.Ja\and
42.79.Ci\and
42.25.Gy}
\end{abstract}

%% file: intro.tex
\section{Introduction}
\label{intro}

It is well-known that a beam of light transmitted through or reflected from a dielectric interface undergoes
a \poldep{} shift of its spatial intensity distribution. The so-called Goos-Hänchen (GH)
\cite{goos_ein_1947,artmann_berechnung_1948}
effect amounts
to a longitudinal shift, i.e.\ a displacement in the plane of incidence, while
the Imbert-Fedorov (IF) shift
\cite{imbert_calculation_1972}
can be observed transverse to this plane.
These positional shifts are connected with angular counterparts
\cite{merano_observing_2009,aiello_duality_2009}.
Both, the GH
\cite{bretenaker_direct_1992,emile_measurement_1995,jost_observation_1998,bonnet_measurement_2001,merano_observation_2007}
and the IF shift
\cite{pillon_experimental_2004,dasgupta_experimental_2006,menzel_imbert-fedorov_2008,mnard_imagingspin_2009}
have been verified experimentally in a number of configurations
while also the theoretical understanding of those effects has advanced
significantly
\cite{berman_goos-hnchen_2002,aiello_role_2008,bliokh_goos-hnchen_2009}.

The IF shift is also known as the Spin Hall Effect of Light (SHEL)
\cite{onoda_hall_2004,bliokh_conservation_2006,hosten_observation_2008}
due to its resemblance to the
Spin Hall Effect
in solid state physics.
It amounts to
a displacement of a circularly polarized beam perpendicular to the plane of incidence,
where the direction depends on the beam's helicity or photon spin.
Consequently, a linearly polarized beam will split into components of different helicity.

The geometric Spin Hall Effect of Light
\cite{aiello_transverse_2009}
is a novel phenomenon, which like SHEL amounts to a 
spin-\dep{}
shift or split of the intensity distribution
of an obliquely incident light beam.
This effect depends significantly on the geometric properties of the detection
system
and, beyond the detection process, no light-matter interaction  is required.

This article is structured as follows:
In the following section the original geometric SHEL is reviewed and
results not explicitly given in \cite{aiello_transverse_2009} are
provided.
In sections \ref{sec:TiltedPol} and \ref{sec:TiltedPolExp}, we introduce a theoretical model for an
arbitrarily oriented planar polarizing interface
\cite{fainman_polarization_1984,aiello_nonparaxial_2009} and provide a scheme for its experimental realization.
Finally, we find that a light beam crossing a tilted polarizing interface undergoes a shift twice as large as the one found in the case of a tilted reference frame.
Therefore, the methods presented in this article lead to a straightforward measurement of the geometric Spin Hall Effect of Light.

%% file: TiltedField.tex
\section{Geometric Shift in a Tilted Reference Frame}
\label{sec:TiltedField}

\begin{figure}
\includegraphics[width=\columnwidth]{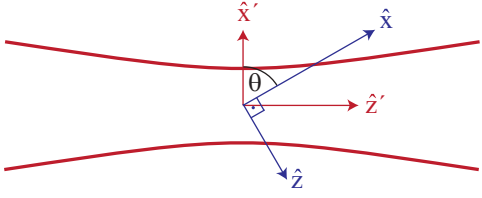}
	\caption{Geometry of the problem: A Gaussian laser beam (red)
propagating in direction of $\vh z'$
	is observed in a plane ($\vh x$, $\vh y$) tilted with
respect to $\vh z'$. The direction $\vh y = \vh y'$ is perpendicular to the
drawing plane.}
	\label{fig:TiltedRF}
\end{figure}

The geometric Spin Hall Effect of Light
\cite{aiello_transverse_2009} occurs when a
circularly polarized beam of light is observed 
in a plane not perpendicular to its direction of
propagation.
This effect amounts to a spatial shift of the intensity distribution
with the intensity being defined as the flux of the Poynting vector
through the detector plane.
We stress that the geometric SHEL only depends on the geometry of the
setup
(\figref{TiltedRF}), the state of polarization and the effective response
function of the detector.

\begin{figure*}
\includegraphics[width=\textwidth]{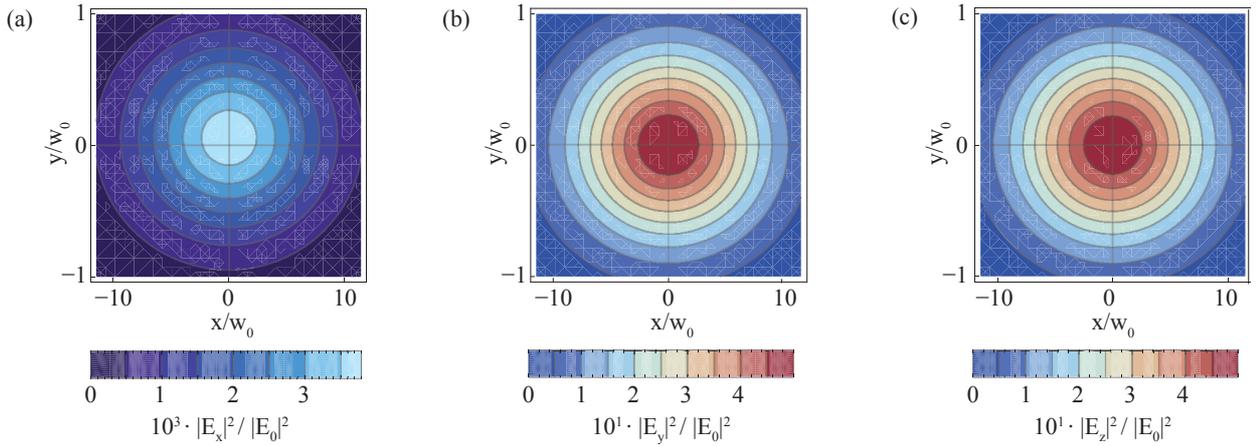}
	\caption{Electric field energy density distribution of a Gaussian light
beam (circular polarization) impinging obliquely ($\theta = \angle(\vh k, \vh z)
= 85\degree$) on a detector.
The components
$|E_x\bvrp|^2$,
$|E_y\bvrp|^2$,
and $|E_z\bvrp|^2$
in the detector reference frame ($\vh x$, $\vh y$, $\vh z$) are shown on
color scales. $|E_0|^2$ is a common normalization constant and $w_0$ is the beam waist.
(a) $|E_x\bvrp|^2$ is clearly shifted in the positive $\vh y$ direction.
(b) $|E_y\bvrp|^2$ exhibits no such shift.
(c) While not visible in this pictorial representation, $|E_z\bvrp|^2$ is shifted in the negative
$\vh y$ direction. Note that the relative weight $w_z$ of this component is more than two orders of magnitude larger than
$w_x$.
}
	\label{fig:TiltedField}
\end{figure*}

Here we give explicit results for
a fundamental Gaussian light beam traveling in direction of
$\vh k = \sin\bt\,\vh x + \cos\bt\,\vh z$ detected in the plane ($\vh x$, $\vh
y$).
The normal $\vh z$ to the detector surface and the propagation vector $\vh k\nparallel\vh z$
unambiguously define a plane of incidence.
As shown by \AAetal{} 
\cite{aiello_transverse_2009},
the intensity barycenter or centroid of a circularly polarized beam
is shifted in direction of $\vh y$ perpendicular to the plane of incidence.
This displacement is equal to
\begin{equation}
	\label{eq:yPoynting}
	\Sbarycenter =
	\frac{\lambda}{4\pi}\,\sigma\,\tan\bt\textrm.
\end{equation}
The superscript $\vec S$ indicates that the centroid was evaluated with respect
to the Poynting vector flux while the subscript TD refers to a tilted
detector.
The shift depends on the helicity $\sigma = \pm 1$ (for left or right hand
circular polarization) and is prominent at grazing incidence $\theta
\rightarrow 90\degree$
where it amounts to a displacement larger than the wavelength $\lambda$.

In \cite{aiello_transverse_2009} it was noted that the energy density (ED) distribution
exhibits no such effect
\begin{math}
\left(
	\EDbarycenter = 0
\right)
\end{math}, which underlines the
dependence on the detector response.
The apparent discrepancy can be understood by decomposing the electric field
energy density
\begin{equation}
\label{eq:ubvrp}
u\bvrp = |\vec E\bvrp|^2 = \Sum_{l=x,y,z} |E_l\bvrp|^2
\end{equation}
into terms depending on one Cartesian component of the electric field in the
detector reference frame ($\vh x$, $\vh y$, $\vh z$) only.
$u\bvrp$ is a distribution in the observation plane and 
$\vec r_\perp = x\,\vh x+y\,\vh y$ is a two-\dimensional{} position vector.
This decomposition is depicted in \figref{TiltedField}.

\newcommand{\SumWlDl}{\Sum_{l={x,y,z}} w_l\,\Delta_l} 

Analogously to \eqref{eq:ubvrp}, we decompose the barycenter $\EDbarycenter$ of
the energy density as
\begin{align}
	\label{eq:EDbary}
	\EDbarycenter &= \frac{\iint y |\vec E\bvrp|^2\dif x\dif y}
	{\iint |\vec E\bvrp|^2\dif x\dif y}
	=
	\SumWlDl\textrm,
\end{align}
where we define
\begin{equation}
\label{eq:wl}
w_l := \frac{\iint |E_l\bvrp|^2 \dxy}{\iint |\vec E\bvrp|^2 \dxy}
\end{equation}
as the relative weight of the field component $|E_l|^2$ and
\begin{equation}
\label{eq:Deltal}
\Delta_l = \frac{\iint y|E_l\bvrp|^2 \dxy}{\iint |E_l\bvrp|^2 \dxy} 
\end{equation}
as the contribution of this component to the total shift.
From a straightforward application of equations
\eqref{eq:wl} and \eqref{eq:Deltal}
to the electric field distribution $\vec E\bvr$ of a
circularly polarized
Gaussian light beam
one finds
\begin{subequations}
\label{eq:Delta}
\begin{align}
	\label{eq:Deltax}
	\Delta_x &=
	\frac{\lambda\,\sigma\,\tan\bt}{2\pi}
	+\gO(\theta_0^2)\textrm,\\
	\label{eq:Deltay}
	\Delta_y &=
	0
	+\gO(\theta_0^2)\textrm,\\
	\label{eq:Deltaz}
	\Delta_z &=
	- \frac{\lambda\,\sigma\,\cot\bt}{2\pi}
	+\gO(\theta_0^2)\textrm,
\end{align}
\end{subequations}
and within the same approximation
\newcommand{\divergence}{\ensuremath{\theta_0 = 2/(k\,w_0) = \lambda/(\pi\,w_0)}}
\begin{subequations}
\label{eq:weight}
\begin{align}
	w_x &= \frac12\cos^2\bt\textrm,\\
	w_y &= \frac12\textrm,\\
	w_z &= \frac12\sin^2\bt\textrm{,}
\end{align}
\end{subequations}
where \divergence{} is the angular divergence of the beam
\cite{mandel_optical_1995}.

Substituting equations \eqref{eq:Delta} and \eqref{eq:weight} into \eqref{eq:EDbary}
we verify that the barycenter of a light beam's energy density 
\begin{equation}
	\label{eq:yED}
	\EDbarycenter = \SumWlDl =  0
\end{equation}
does not shift under rotation of the reference frame.
This result underlines the scalar nature of the energy density.

We remind the reader that the Poynting vector flux through the detector surface
\begin{align}
	\label{eq:s_z}
	s_z\bvrp = \vec s\bvrp\cdot\vh z \propto
	\left[\vec E\bvrp\times\vec B^\ast\bvrp\right]\cdot\vh z\textrm,
\end{align}
where $\vh z$ is the surface normal,
is a distribution different from $u\bvrp$ and exhibits a net shift
\begin{equation}
	\Sbarycenter \propto
	\Delta_x
	\textrm,
\end{equation}
where $\Delta_x$ is given in \eqref{eq:Deltax}.

This connects the geometric SHEL with the fundamental question about
the local response of a po\-si\-tion-sen\-si\-tive detector.
The definition of the Poynting vector flux as the intensity is
motivated by Poynting's theorem.
However, this choice is debatable
as the theorem does not define the local
Poynting vector unambiguously \cite{berry_optical_2009}
and the definition given in equation \eqref{eq:s_z} depends on the state of polarization.
Contrarily, the response function of a real \pol-in\dep{} detector is more likely to
be isotropic, i.e.\ to depend on
$u = |E_x|^2 +|E_y|^2 + |E_z|^2$,
and thus yield
$\EDbarycenter = 0$.

The shift $\Delta_x$ can be detected directly if
a detection
scheme is used where a plane of observation ($\vh x$, $\vh y$) can be chosen arbitrarily
and the effective response function depends on $|E_x|^2$ but not on
$|E_z|^2$.
The \poldep{} absorption in semiconductor quantum wells
\cite{weiner_strong_1985,weiner_highly_1985,kihara_rurimo_usingquantum_2006}
or
single molecules
\cite{sick_orientational_2000,novotny_longitudinal_2001}
can in principle be used to build a suitable detector.

In the
remaining part of this article we develop an alternative strategy to measure the characteristic shift caused by
the geometric SHEL.

%% file: TiltedPol.tex
\section{The Ideal Polarizer Model}
\label{sec:TiltedPol}

The operation performed by a polarizing optical element is commonly only
determined for
normally incident beams which can be approximated as planar wave fronts.
In this case the action of
a polarizer is described as a projection
within a plane perpendicular to the direction of propagation $\vh k
= \frac1{|\vec k|}\,\vec k$.
This simple model fails to describe the operation of a polarizing element
when no such assumptions about the light field are made, as it is the case in this
article where we deal with obliquely incident beams.

To overcome this limitation, Fainman and Shamir (FS) have proposed a model
describing an arbitrarily oriented ideal polarizer
\cite{fainman_polarization_1984}.
They introduce a
three-\dimensional{} complex-valued unit vector $\vh P$ and describe
the action of a polarizer as follows:
The electric field vector $\vec E(\vec k)$ of each
plane wave of the incident beam's angular spectrum is projected onto \eP,
a unit vector perpendicular to $\vec k$:
\newcommand{\FSden}{\sqrt{1-\left|\vh k\cdot\vh P\right|^2}}
\begin{align}
\label{eq:e_P}
\eP(\vh k) &= 
\frac{\vh k\times(\vh P \times\vh k)}{\FSden}
=
\frac{\vh P - \vh k(\vh k\cdot\vh P)}
{\FSden}\\
\label{eq:FSpol}
\vec E(\vec k) &\rightarrow \eP(\vh k)\,
\left[\eP^\ast(\vh k)\cdot\vec E(\vec k)\right]
\end{align}
The model is constructed such that the operation is
idempotent
and does not
change $\vec k$ (\figref[a]{TiltedPol}).

\begin{figure}
	\includegraphics[width=\columnwidth]{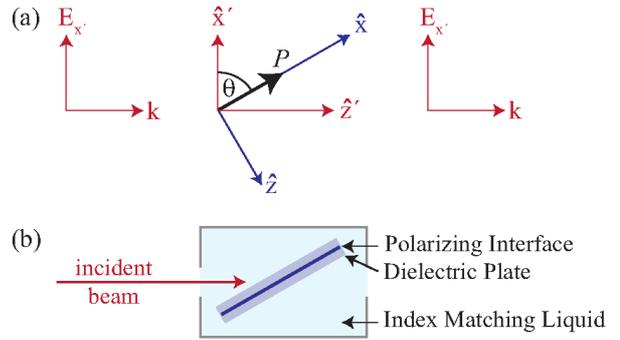}
	\caption{
(a) Interaction of a plane wave propagating in direction of $\vh z'$ with a
tilted polarizer described by $\vh P = \vh x = \cos\bt\,\vh x' + \sin\bt\,\vh
z'$.
Before and after passing the polarizing element the electric field is
perpendicular to $\vh k$.
(b) Thin film polarizer submerged in a tank of index matching liquid. This
scheme allows
to study a tilted polarizing interface eliminating effects of physical boundaries.
	}
	\label{fig:TiltedPol}
\end{figure}

A remarkable characteristic of the FS polarizer is that rotation (around $\vh y$)
has no effect on a single plane wave if $\vh P$ is parallel or perpendicular
to $\vh y$.
Hence, an ideal polarizing element cannot be used to cause an imbalance between the
weights $w_x$ and $w_z$ of the electric field component parallel and perpendicular to the polarizer surface.

However, as we will show in \secref{PolShift}, the rotation gives rise to
an effect on bounded beams similar to the geometric SHEL in a tilted reference
frame.
In the following section we will propose an experimental
realization of an arbitrarily oriented polarizing interface.

%% file: TiltedPolExp.tex
\section{Experimental realization of a universal polarizing interface}
\label{sec:TiltedPolExp}

We propose a scheme 
using only off-the-shelf optical components
to study the interaction of a light beam with a polarizing element in any
geometry
(\figref[b]{TiltedPol}).
For this purpose we model a real polarizer as a composite device consisting of
an infinitely thin polarizing interface sandwiched between dielectric plates
with a refractive index $n>1$.
Since commercial thin film polarizers
are typically protected from the environment by either a substrate or a coating
on each face, our model is close to the realistic scenario.

The interaction of the light field with an air-dielectric boundary is
\poldep{} and changes the direction of propagation $\vh k$ of a plane wave.
Those well-known effects, described by Snell's law of refraction and Fresnel's formulas
\cite{hecht_optik_2005},
are caused by the refractive index step.
At grazing incidence refraction is so severe that inside the front dielectric
plate
the 
angle between the direction of propagation 
and the surface normal is always significantly smaller than the corresponding angle $\theta$ in air.
Therefore, it is desirable to
eliminate the change of the refractive index at the physical
boundary.
This can be done, for example, by embedding
the glass-polarizer-glass system
in an index-matched environment.
As a side effect this also eliminates
unwanted
Imbert-Fedorov and Goos-Hänchen shifts.

It is not common for vendors to specify the behavior of
polarization optics
under non-normal incidence.
Measurements in our laboratory using the proposed scheme indicate that the FS
model is suitable to describe a tilted polarizing interface.
A detailed
investigation
will be reported elsewhere.

%% file: PolShift.tex
\section{Geometric Shift at a Polarizing Interface}
\label{sec:PolShift}

In \secref{TiltedField},
we described the geometric SHEL as an effect which occurs
for a light beam
in vacuum
when the plane of observation is tilted with respect to the direction of
propagation.
The predicted shift depends on
specific
assumptions about the detection
process
and, therefore, cannot be easily verified.
In this section, we shall show that an ideal polarizer performs an operation on a
light beam that amounts to the characteristic
geometric SHEL
shift independent of the
properties of the detection system.

\begin{figure*}
\includegraphics[width=\textwidth]{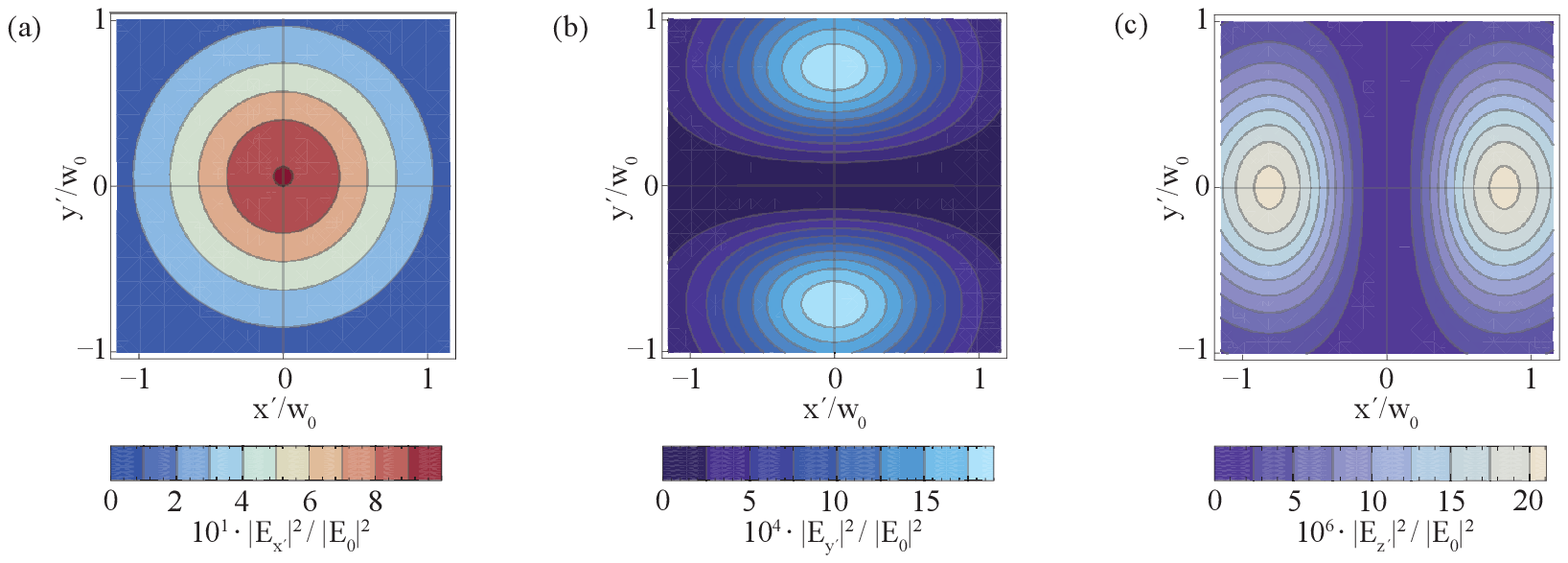}
	\caption{Electric field energy density distribution of a Gaussian light
beam (circular polarization) after interacting with a tilted polarizer
$\left(\vh P = \cos(85\degree)\,\vh x' + \sin(85\degree)\,\vh z'\right)$.
The components
$|E_{x'}\bvrp|^2$,
$|E_{y'}\bvrp|^2$,
and $|E_{z'}\bvrp|^2$
in the beam's natural reference frame ($\vh x'$, $\vh y'$, $\vh z'$) are shown on
color scales. $|E_0|^2$ is a common normalization constant and $w_0$ is the beam waist.
(a) $|E_{x'}\bvrp|^2$ is shifted as in \figref[a]{TiltedField}.
(b), (c) $|E_{y'}\bvrp|^2$
and $|E_{z'}\bvrp|^2$
are not shifted and their relative weights are negligible.
}
\label{fig:PolShift}
\end{figure*}

As in the case of the tilted detector, we assume the incident beam 
to travel in direction of $\vh k =: \vh z'$ and to
have a fundamental Gaussian
profile with its barycenter at 
$\lrangle{x'} = 0$ and $\lrangle{y'}=0$,
where ($\vh x'$, $\vh y'$, $\vh z'$) is the beam's natural reference frame and
$\vh y'$ coincides with $\vh y$
(geometry as in \figref{TiltedPol}).
The polarizer shall be oriented along
\begin{equation}
 \label{eq:P=x}
  \vh P = \vh x = \cos\bt\,\vh x' + \sin\bt\,\vh z'
  \textrm{.}
\end{equation}

Using dimensionless coordinates
$\tx = x'/w_0$,
$\ty = y'/w_0$,
$\tz = z'/L$
and
$\vec r' = \bxez$
where
$w_0$ denotes the beam waist and
$L = k\,w_0^2/2$ the Rayleigh length, the fundamental solution
of the paraxial scalar wave equation is:
\newcommand{\psiden}{{1+i\tz}}
\begin{equation}
	\psi(\vec r')
	=
	\frac{1}{\psiden}\,\exp\left(
	-\frac{\tx^2 +\ty^2}\psiden
	\right)
\end{equation}
Let
$\vh u = \frac1{\sqrt2}\,\left(\vh x' \pm i\, \vh y'\right)$ be a complex unit vector
denoting left or right hand circular states of polarization.
The electric field of a fundamental Gaussian beam
can be
written as
\begin{equation}
	\label{eq:EGaussBeam}
	\vec E(\vec r') \propto
	\exp\left(
	\frac{i\,2\tz}{\theta_0^2}
	\right)
	\left(
	\vh u
	 + \frac{i\,\theta_0}2\,
	\vh z\,(\vh u \cdot \nabla_\perp)
	\right)\psi(\vec r')
	\textrm{,}
\end{equation}
where 
$\divergence$ is the angular spread of the beam
and
$\nabla_\perp = (\partial_{\tx}, \partial_{\ty})$
is the transverse gradient
operator
\cite{haus_photon_1993}.

To apply the FS polarizer model \eqref{eq:FSpol},
the electric field \eqref{eq:EGaussBeam} must be expressed in its angular
spectrum representation $\vec E\bvk$
and the beam after interacting with the tilted polarizing element becomes:
\begin{equation}
	\label{eq:EafterPol}
	\vec E(\vec r') = \iiint\exp(i\,\vec k\cdot\vec r')\eP(\vh k)\,
	\left[\eP^\ast(\vh k)\cdot\vec E(\vec k)\right]\dif^3\vec k
\end{equation}

From the electric field distribution \eqref{eq:EafterPol},
we calculate the energy density of a light beam after passing through a tilted polarizer.
The Cartesian components thereof are
depicted in \figref{PolShift}.
Unlike in \secref{TiltedField}, in
this case the evaluation is performed in the beam reference frame ($\vh x'$,
$\vh y'$, $\vh z'$) where $\vh z' = \vh k$.
Decomposing the energy density barycenter \EDbarycenter[TP] as in \eqref{eq:EDbary} one finds 
\begin{subequations}
\label{eq:Deltap}
\begin{align}
	\label{eq:Deltaxp}
	\frac{\Delta_{x'}}{w_0}
	&= \frac{\theta_0}{2}\,\sigma\,\tan\bt +
\gO(\theta_0^2)\textrm,\\
\displaybreak[0]
	\frac{\Delta_{y'}}{w_0} &= 0+ \gO(\theta_0^2)
\textrm{, and}\\
\displaybreak[0]
	\frac{\Delta_{z'}}{w_0} &= 0+ \gO(\theta_0^2)\textrm,
\end{align}
\end{subequations}
where $\sigma = \pm1$ is the helicity of the beam.
Since we observe a collimated light beam
in its natural reference frame
after passing through a linear polarizer, the weights
\begin{subequations}
\label{eq:weightsp}
\begin{align}
	w_{y'} &= 0 + \gO(\theta_0^2)\textrm{ and } \\
	w_{z'} &= 0 + \gO(\theta_0^2)
\end{align}
vanish, and, consequently,
\begin{align}
 	w_{x'} &= 1 + \gO(\theta_0^2)\textrm.
\end{align}
\end{subequations}

The centroid of a circularly polarized beam
transmitted across a tilted polarizer
is
thus
\begin{align}
	\label{eq:yTP}
	\EDbarycenter[TP] &= \Sbarycenter[TP] = \Delta_x
	= \frac{\lambda}{2\pi}\,\sigma\tan(\theta)
	+ \gO(\theta_0^2)
\end{align}
and can be measured with any detector sensitive to any weighted sum of
$|E_{x'}|^2$, $|E_{y'}|^2$, and $|E_{z'}|^2$ if the weight of the first term
does not
vanish. Standard detectors such as photodiodes and CCD cameras certainly
meet this requirement.

We stress that equation
\eqref{eq:yTP} resembles the original result \eqref{eq:yPoynting}.
The displacement introduced by the tilted polarizer
\begin{equation}
	\Sbarycenter[TP] = 2\Sbarycenter
\end{equation}
is twice the one found for the Poynting vector flux
through a tilted plane of observation.
Furthermore, this article gives a straightforward recipe to measure the shift.

Since equation (5) from \cite{aiello_transverse_2009} is generally valid, both
shifts are connected to a transverse angular momentum,
which occurs when the angular momentum calculated in the local frame attached to the light beam
is projected upon a global frame tilted with respect to the former.
For the geometric SHEL (occurring at a tilted detector), the projection is implicitly given by the definition of intensity
as the flux $S_z = \vec S\cdot \vh z$ of the Poynting vector across the detector surface.
Conversely, in the case of the tilted polarizing interface, as described in this section,
the projection is caused by the polarizer.
Therefore, we can conclude that both shifts arise because of the projection of the
intrinsic longitudinal angular momentum of a circularly polarized light beam
onto a tilted reference frame.
We remind the reader that 
beyond those geometric projections, no physical interaction occurs.

%% file: conclusion.tex
\section{Conclusion}

First, the original geometric Spin Hall Effect of Light,
as described by \AAetal{},
was illustrated using an
explicit decomposition of a light beam's energy density
in a tilted reference frame. We showed that in order to observe this effect,
which occurs in
vacuum and amounts to a \poldep{} shift, a suitably tailored detection system is
required.

Then, a novel type of geometric SHEL
occurring at a polarizing interface
was introduced.
To this end, we discussed
a theoretical model for an ideal polarizer
and suggested an
experimental implementation thereof.
The light field of a collimated laser beam
transmitted across
such a polarizer was evaluated.
In the case of the polarizing interface being tilted with respect to the direction of propagation,
a beam displacement resembling the original geometric SHEL was found.
This shift does not depend on the detection process and
can be measured in a straightforward way by using the scheme proposed in this article.

The effect derived in our work is unavoidable when a circularly polarized
light beam passes through a polarizing interface tilted with respect to the direction of propagation.
This underlines the importance of the geometric SHEL 
as polarization is a fundamental property of the light field and numerous optical devices are \poldep.

%% file: ack.tex
\section*{Acknowledgements}

A.A.\ acknowledges support from the Alexander von Humboldt foundation.